\documentclass[preprint2,twoside]{hwo}

\usepackage{lipsum}
\usepackage{threeparttable}  

\newcommand{\oiii}{O\,{\sc iii}}

\newcommand{\siiv}{Si\,{\sc iv}}

\newcommand{\cii}{[C\,{\sc ii}]}
\newcommand{\ciii}{C\,{\sc iii}}
\newcommand{\civ}{C\,{\sc iv}}

\def\ly{$\lambda$}

\def\cii{C\,{\sc ii}}
\def\ciii{C\,{\sc iii}}
\def\civ{C\,{\sc iv}}

\def\niii{N\,{\sc iii}}
\def\niv{N\,{\sc iv}}
\def\nv{N\,{\sc v}}
\def\oi{O\,{\sc i}}
\def\oiii{O\,{\sc iii}}

\def\siiv{Si\,{\sc iv}}
\def\Siii{Si\,{\sc ii}}

\def\siiii{Si\,{\sc iii}}

\def\Q0059{Q0059--2735}

\def\S2S3{S2S3}

\def\nh{\ifmmode n_\mathrm{\scriptscriptstyle H} \else $n_\mathrm{\scriptscriptstyle H}$\fi}
\def\ne{\ifmmode n_\mathrm{\scriptstyle e} \else $n_\mathrm{\scriptstyle e}$\fi}
\def\Te{\ifmmode T_\mathrm{\scriptstyle e} \else $T_\mathrm{\scriptstyle e}$\fi}
\def\Qh{\ifmmode Q_\mathrm{\scriptstyle H} \else $Q_\mathrm{\scriptstyle H}$\fi}
\def\Uh{\ifmmode U_\mathrm{\scriptstyle H} \else $U_\mathrm{\scriptstyle H}$\fi}
\def\Nh{\ifmmode N_\mathrm{\scriptstyle H} \else $N_\mathrm{\scriptstyle H}$\fi}
\def\Nhi{\ifmmode N_\mathrm{\scriptstyle HI} \else $N_\mathrm{\scriptstyle HI}$\fi}
\def\Uhhp{\ifmmode U_\mathrm{\scriptstyle H,HP} \else $U_\mathrm{\scriptstyle H,HP}$\fi}
\def\Nhhp{\ifmmode N_\mathrm{\scriptstyle H,HP} \else $N_\mathrm{\scriptstyle H,HP}$\fi}
\def\Uhvhp{\ifmmode U_\mathrm{\scriptstyle H,VHP} \else $U_\mathrm{\scriptstyle H,VHP}$\fi}
\def\Nhvhp{\ifmmode N_\mathrm{\scriptstyle H,VHP} \else $N_\mathrm{\scriptstyle H,VHP}$\fi}
\def\Nion{\ifmmode N_\mathrm{\scriptstyle ion} \else $N_\mathrm{\scriptstyle ion}$\fi}

\def\Zsun{\ifmmode {\rm Z}_{\odot} \else $Z_{\odot}$\fi}
\def\Msun{\ifmmode {\rm M}_{\odot} \else M$_{\odot}$\fi}
\def\kms{\ifmmode {\rm km~s}^{-1} \else km~s$^{-1}$\fi}
\def\Lya{\ifmmode {\rm Ly}\alpha \else Ly$\alpha$\fi}
\def\Lyb{\ifmmode {\rm Ly}\beta \else Ly$\beta$\fi}
\def\Lyg{\ifmmode {\rm Ly}\gamma \else Ly$\gamma$\fi}
\def\Lyd{\ifmmode {\rm Ly}\delta \else Ly$\delta$\fi}
\def\neaod{\ifmmode n_\mathrm{\scriptscriptstyle AOD} \else $n_\mathrm{\scriptscriptstyle AOD}$\fi}
\def\necrit{\ifmmode n_\mathrm{\scriptstyle cr} \else $n_\mathrm{\scriptstyle cr}$\fi}
\def\ncr{\ifmmode n_\mathrm{\scriptstyle cr} \else $n_\mathrm{\scriptstyle cr}$\fi}
\def\nepi{\ifmmode n_\mathrm{\scriptscriptstyle PI} \else $n_\mathrm{\scriptscriptstyle PI}$\fi}
\def\gtorder{\mathrel{\raise.3ex\hbox{$>$}\mkern-14mu\lower0.6ex\hbox{$\sim$}}}
\def\ltorder{\mathrel{\raise.3ex\hbox{$<$}\mkern-14mu\lower0.6ex\hbox{$\sim$}}}

\def\vro{\ifmmode v_\mathrm{\scriptscriptstyle 1, \scriptstyle r} \else $v_\mathrm{\scriptscriptstyle 1, \scriptstyle r}$\fi}
\def\vrc{\ifmmode v_\mathrm{\scriptscriptstyle 2, \scriptstyle r} \else $v_\mathrm{\scriptscriptstyle 2, \scriptstyle r}$\fi}
\def\vzo{\ifmmode v_\mathrm{\scriptscriptstyle 1, \scriptstyle z} \else $v_\mathrm{\scriptscriptstyle 1, \scriptstyle z}$\fi}
\def\vzc{\ifmmode v_\mathrm{\scriptscriptstyle 2, \scriptstyle z} \else $v_\mathrm{\scriptscriptstyle 2, \scriptstyle z}$\fi}

\newcommand{\fescLyC}{\textit{f}$_{\text{esc}}^{\text{ LyC}}$}

\input{hwo.h}

\begin{document}

\title{\textbf{\LARGE Spatially Resolving the Fundamental Elements of Reionization \\in Galaxies}}

\author {\textbf{\large Xinfeng Xu,$^{1,2}$ Stephan McCandliss,$^3$  Allison Strom,$^{1,2}$ Hsiao-Wen Chen,$^4$ Yumi Choi,$^5$ Annalisa Citro,$^6$ Håkon Dahle,$^7$ Matthew J. Hayes,$^8$ Anne Jaskot,$^9$ Logan Jones,$^{10}$ Gagandeep Kaur,$^{11}$ Themiya Nanayakkara,$^{12}$ Alexandra Le Reste,$^6$}}

\affil{$^1$\small\it Department of Physics and Astronomy, Northwestern University,
2145 Sheridan Road, Evanston, IL, 60208, USA}
\affil{$^2$\small\it Center for Interdisciplinary Exploration and Research in
Astrophysics (CIERA), 1800 Sherman Avenue,
Evanston, IL, 60201, USA}
\affil{$^3$\small\it Center for Astrophysical Sciences, Department of Physics \& Astronomy, Johns Hopkins University, Baltimore, MD 21218, USA}
\affil{$^4$\small\it The University of Chicago, Department of Astronomy \& Astrophysics, 5640 S. Ellis Ave., Chicago, IL 60637, USA}
\affil{$^5$\small\it NSF National Optical-Infrared Astronomy Research Laboratory, 950 North Cherry Avenue, Tucson, AZ 85719, USA}
\affil{$^6$\small\it Minnesota Institute for Astrophysics, University of Minnesota, 116 Church Street SE, Minneapolis, MN 55455, USA}
\affil{$^7$\small\it Institute of Theoretical Astrophysics, University of Oslo, P.O. Box 1029, Blindern, NO-0315, Oslo, Norway}
\affil{$^8$\small\it Department of Astronomy and Oskar Klein Centre for Cosmoparticle Physics, Stockholm University, SE-10691, Stockholm, Sweden}
\affil{$^9$\small\it Department of Astronomy, Williams College, Williamstown, MA 01267, USA}
\affil{$^{10}$\small\it Space Telescope Science Institute, 3700 San Martin Drive, Baltimore, MD 21218, USA}
\affil{$^{11}$\small\it Graz University of Technology, 8010 Graz, Austria}
\affil{$^{12}$\small\it Centre for Astrophysics and Supercomputing, Swinburne University of Technology, Hawthorn, Australia}



\begin{abstract}
Cosmic reionization marks a critical epoch when the first galaxies ionized the intergalactic medium through the escape of Lyman continuum (LyC) radiation. Young, massive star clusters are believed to be the primary LyC sources, yet the physical mechanisms enabling LyC escape remain poorly understood. Most existing studies rely on spatially integrated observations, which lack the resolution to resolve internal galaxy structure and pinpoint where and how LyC photons escape. To address this, we propose a science case for the Habitable Worlds Observatory (HWO) that enables spatially resolved spectroscopy of LyC-emitting star clusters and their environments in low-redshift galaxies. This requires a UV integral field unit (IFU) with coverage down to $\sim$ 900\AA\ and a spatial resolution of 10–100 pc—capabilities essential for directly detecting LyC escape and mapping the surrounding interstellar medium. With such instrumentation, we will map cluster-scale LyC escape fractions, characterize the physical conditions of the surrounding interstellar medium, and directly observe feedback-driven outflows that facilitate LyC leakage. These observations will enable novel calibrations of indirect LyC indicators at unprecedented spatial resolution and establish direct connections between local LyC processes and those in high-redshift, clumpy star-forming galaxies. In the long run, this program will build the physical framework needed to understand how galaxies reionized the early universe and shaped its subsequent evolution.
  \\
  \\
\end{abstract}

\vspace{2cm}

\section{Introduction}
\label{sec:intro}
After the Big Bang, the universe expands and cools, leading to a period when it is filled with neutral hydrogen gas. During this time, often called the cosmic dark ages, the universe becomes opaque to most radiation, and light cannot travel freely. However, there is an important transition, known as cosmic reionization, during which the universe becomes ionized again and gradually transparent to light. Understanding how and when reionization occurs directly addresses key questions outlined in the Astro2020 Decadal Survey \citep{Astro2020}, particularly within the Cosmic Ecosystems theme, such as ``How did the intergalactic medium (IGM) and the first sources of radiation evolve from cosmic dawn through the epoch of reionization?''

\begin{figure*}[ht]
\center

	\includegraphics[page = 1,angle=0,trim={0.0cm 0.0cm 30cm 0.0cm},clip=true,width=0.77\linewidth,keepaspectratio]{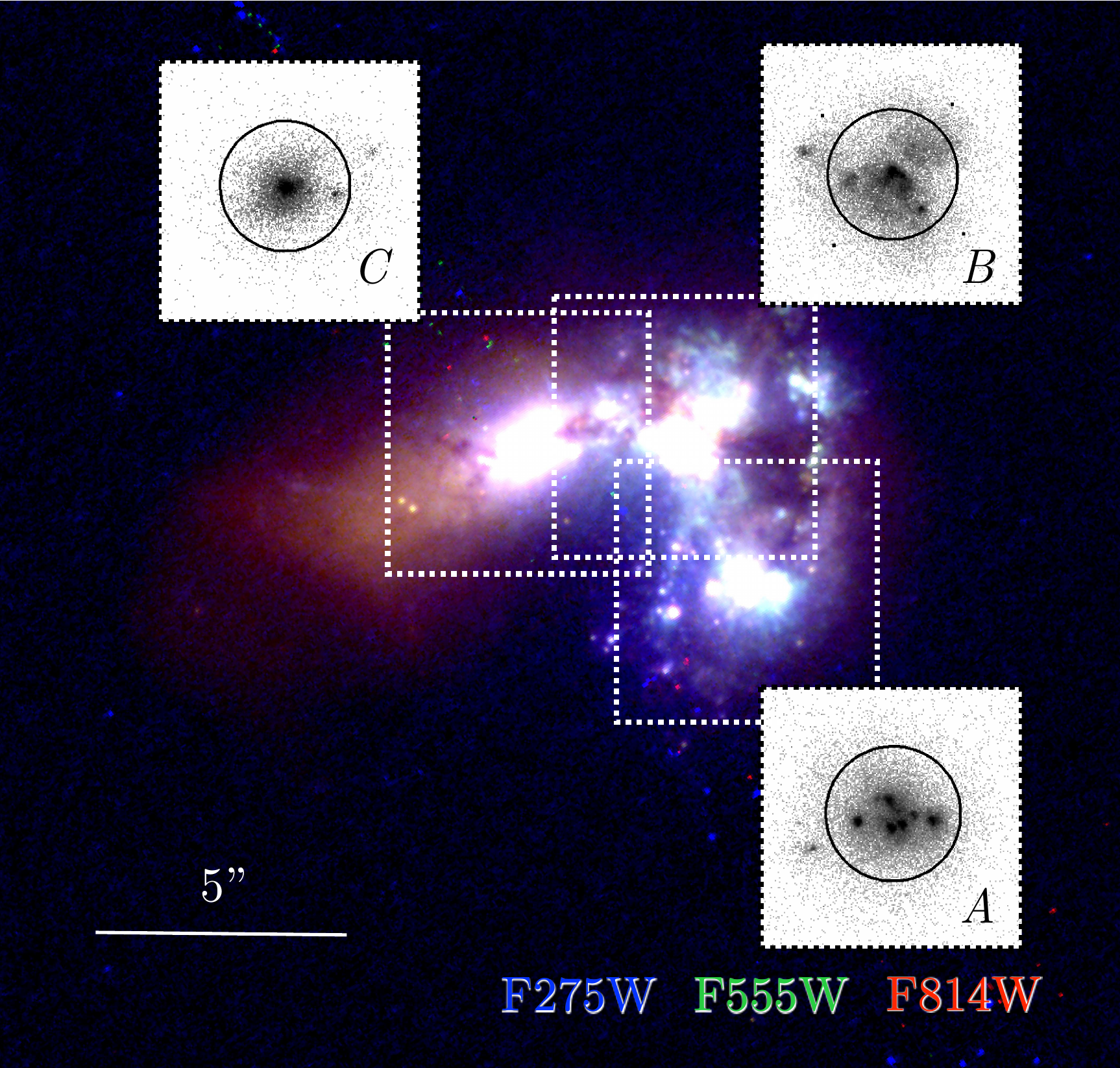}

\caption{\normalfont{Hubble Space Telescope (HST) color-composite images of Haro 11, a nearby ($z=0.02$) low-mass starburst galaxy emitting Lyman continuum (LyC) photons \citep[figure taken from H. Chen et al. in prep., see also][]{Komarova24}. Insets show near-UV HST/COS images of star clusters A, B, and C, with black circles denoting the 2.5\arcsec\ HST/COS aperture. Clusters B and C exhibit LyC leakage, with escape fractions of $f_{\rm esc}=(3.4\pm 2.9)$\% and ($5.1\pm 4.3$)\%, respectively \citep[][]{Komarova24}, whereas cluster A shows no LyC detection, implying a $2\sigma$ upper limit of $f_{\rm esc}<10$\%. Despite Haro 11’s proximity, resolving individual star clusters on 10–100 pc scales requires angular resolutions of $\sim$ 0.02\arcsec -- 0.2\arcsec.}  }
\label{fig:haro11}
\end{figure*}

The main drivers of cosmic reionization are believed to be the early generations of young stars and galaxies. Stars start to form in galaxies in giant molecular cloud complexes that collapse under the force of gravity and fragments.  The most massive fragments become massive stars that use up their nuclear fuel quickly in a few million years, and once the fuel is exhausted, they explode as supernovae.  During their lifetime, they burn so hot and intensely that the outer layers of these stars are driven away in the form of a wind.  The stellar wind and eventual supernova disrupt the natal cloud, allowing a direct glimpse of the star’s very energetic radiation with E $>$ 13.6 eV—commonly termed ionizing radiation or Lyman continuum (LyC) radiation.  The radiation is so energetic that it rips the surrounding neutral hydrogen gas from which the star formed into its constituent electrons and protons, a process known as ionization. When this light escapes galaxies into the intergalactic space, it causes the ionization of most of the universe’s neutral hydrogen. However, detecting this potent energy proves immensely challenging in distant, high-redshift galaxies (z $\gtrsim$ 4), where the predominantly neutral universe absorbs any escaped photons before they reach Earth.

Nonetheless, if we look at nearby young galaxies with a large enough telescope and sensitivity to very short wavelength far-ultraviolet radiation, we can directly examine the regions within a galaxy that allow the ionizing radiation to escape into the IGM and sustain the universe in it mostly ionized state. These regions are called ‘star clusters’ (SCs), and they represent the most prominent mode of star formation across cosmic history \citep{Adamo20}. When star clusters are young and massive, they are capable of producing significant amounts of LyC photons. If there are preferred structures in these galaxies (e.g., holes in dense clouds caused by galactic winds, see Figure \ref{fig:sc}), some of these LyC radiations will leak to the IGM. Thus, these star clusters that leak LyC (hereafter, LyC clusters) are the fundamental elements in galaxies responsible for cosmic reionization. Understanding the properties of LyC clusters is also crucial to properly answering many other important questions in astrophysics. Strong stellar feedback from young stars in massive clusters (in the form of wind, radiation pressure, and supernova explosions) are known to regulate star formation and drive outflows on a galactic scale \citep[e.g.,][]{Walch15}. The extremely high stellar densities in the cores of massive clusters also cause stellar collisions and merging, a potential channel for seeding the formation of supermassive black holes \citep{Davies11}.

Direct detections of LyC clusters are extremely rare, with only a handful of examples currently known, e.g., Haro 11 (z = 0.02), Sunburst (z = 2.37), and Sunrise arc (z = 6.0) \citep[see, e.g.,][]{Kim23, Pascale23, Vanzella22, Vanzella23, Komarova24}. Even in Haro 11 (Figure \ref{fig:haro11}), despite its proximity, resolving individual clusters on 10–100 pc scales demands angular resolutions of $\sim$ 0.02\arcsec -- 0.2\arcsec. For more distant sources, studies must exploit strong gravitational lensing to reach comparable spatial resolution, since no current facility can directly resolve $\sim$10–100 pc LyC clusters at $\sim$900 \AA. In addition, \cite{Choi20} probed the LyC escape fraction in a nearby ($\sim$3 Mpc) starburst galaxy by analyzing individual stars; yet, in the absence of instruments capable of directly detecting escaping LyC photons, this approach relied on indirect proxies. Thus, our understanding of LyC clusters and their role in driving reionization on galactic scales remains limited, underscoring the need for spatially resolved observations from unbiased surveys.

Two recent surveys have advanced our understanding of LyC escape in star-forming galaxies—but have also underscored the crucial role of spatial resolution. The Low‑Redshift Lyman Continuum Survey (LzLCS) used Hubble Space Telescope (HST)/Cosmic Origin Spectrograph (COS) to measure LyC escape from 66 galaxies, detecting LyC in $\sim$50\% of the cases \citep{Flury22a, Flury22b}. But the observations lack spatial resolution due to the 2.5\arcsec\ COS aperture covering the entire galaxy, obscuring the specific regions driving leakage. To overcome this, the 119-orbit HST imaging program LaCOS targets 41 LzLCS galaxies with multi-filter observations. Early LaCOS results reveal tight correlations between global escape fractions and spatial distributions of UV-bright clusters—demonstrating that sub-parsec spatial resolution is essential to isolate LyC-emitting regions and identify the physical conditions enabling escape \citep{LeReste25}. However, even these deep observations cannot resolve LyC clusters directly or diagnose leakage geometries, underscoring the need for HWO’s unique capability to achieve the required spatial resolution, spectral coverage, and sensitivity.

In this paper, we propose to have the first-ever UV IFU instrument on the HWO to tackle the above challenges. We present the key science objectives for us to understand the properties of SCs and their leaked LyC in Section \ref{sec:objective}. Then we discuss the necessary physical parameters to measure in Section \ref{sec:params}. Finally, we describe the requred observations to reach our goals in Section \ref{sec:obs}.

\section{Science Objectives}
\label{sec:objective}

The major objective is to directly resolve both the clusters themselves and the surrounding interstellar medium (ISM). This is because galaxies exhibit complex internal variations in gas density, metallicity, star formation activity, and feedback processes, which cannot be captured by a single integrated value. While recent studies \citep{Flury22a, Flury22b, Choustikov24, Jaskot24a, Jaskot24b} have developed indirect methods to predict galaxy-integrated LyC escape fractions (\fescLyC) based on global properties, interpreting these measurements remains extremely challenging. Thus, with the necessary capabilities, HWO can revolutionize our understanding of LyC radiation from galaxies by spatially resolving the LyC clusters and their local environments.
We summarize the detailed science objectives below.

First, we need to spatially resolve the LyC clusters at the wavelength range from rest-UV to at least the Lyman edge at 912\AA\ in low-z SF galaxies. This is to directly detect: 1) the escaping LyC radiation from the cluster and 2) the stellar properties of the cluster. The latter will be used to estimate the total amounts of LyC radiation produced from the cluster. Combining both, we can calculate \fescLyC, which is the key property to measure to quantify the effects of LyC clusters to the ionization on interstellar to intergalactic scales. Importantly, low-redshift SF galaxies provide our best opportunity to directly detect escaping LyC radiation, because at higher redshifts (z $\gtrsim$ 4), the neutral IGM absorbs any escaping LyC photons before they reach Earth. By building a well-characterized sample of LyC clusters in nearby galaxies, we can develop and calibrate indirect indicators of LyC leakage, which will be essential for interpreting observations of galaxies during the reionization epoch.

\begin{figure*}[h]
\center

	\includegraphics[page = 1,angle=0,trim={0.0cm 0.0cm 0.0cm 0.0cm},clip=true,width=0.9\linewidth,keepaspectratio]{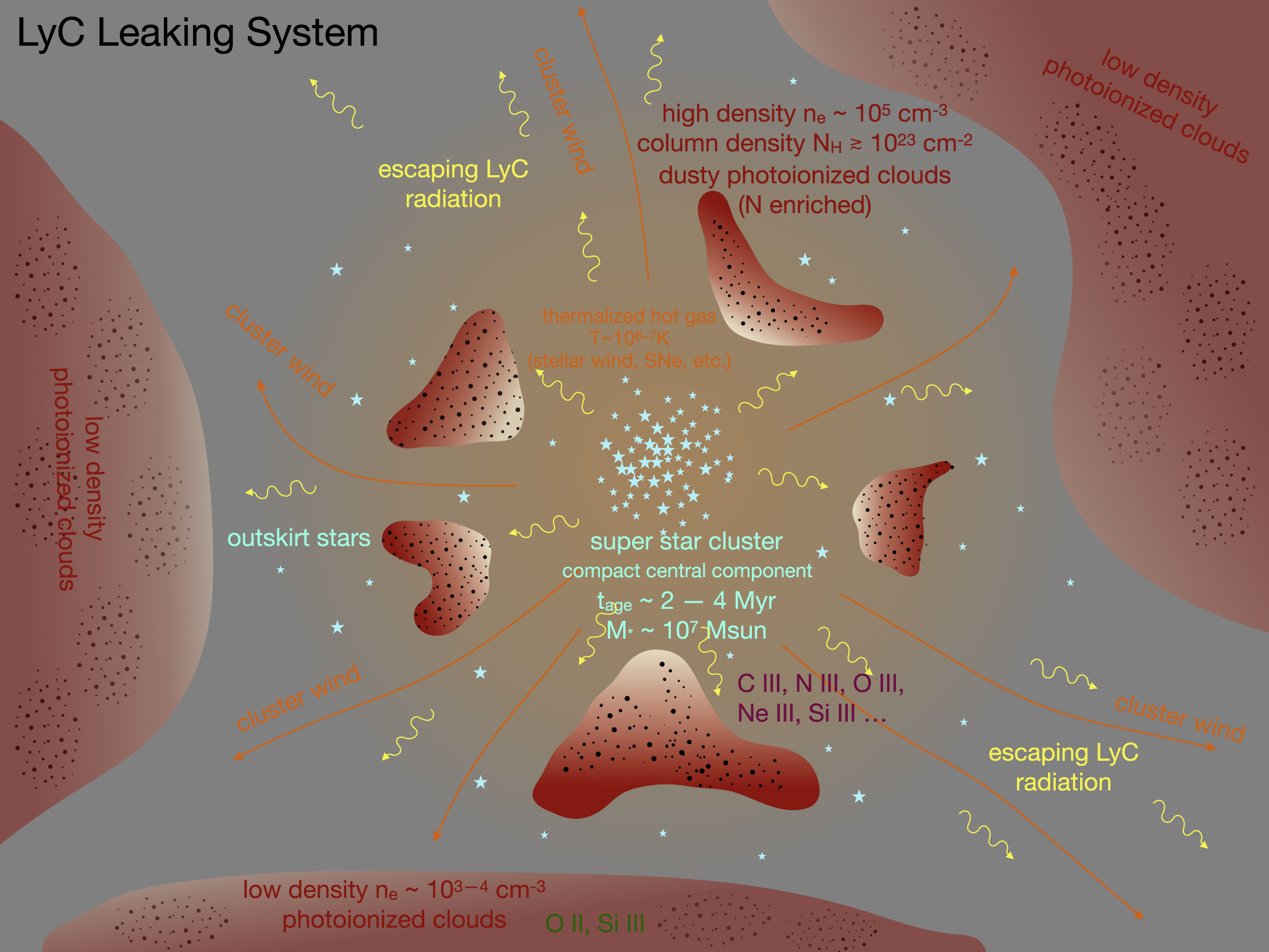}

\caption{\normalfont{A cartoon illustrating the general physical picture for the LyC cluster and the material around it \citep[figure and content taken from][]{Pascale23}. A young super star cluster ($<$ 4 Myr) generates a hot (10$^{6}$–10$^{7}$ K), supersonic wind driven by stellar winds and early supernovae. Dense, high-pressure clouds (\ne\ $\sim$ 10$^{5}$ cm$^{-3}$) enriched by stellar ejecta form near the cluster ($\sim$ 5–10 pc), emitting strong high-ionization lines. Further out, lower-density clouds produce low-ionization lines. The surrounding medium is porous, allowing some LyC radiation to escape through low-density channels.}  }
\label{fig:sc}
\end{figure*}

Secondly, we aim to spatially resolve the interstellar medium (ISM) clouds surrounding LyC clusters (red regions in Figure \ref{fig:sc}) to study the physical conditions of the ionized gas illuminated by these clusters. Key properties to measure include gas-phase abundances, electron density and temperature, pressure, and dust extinction. Obtaining these measurements across a representative sample of LyC clusters will allow us to compare local ISM environments to those in early galaxies (z $\gtrsim$ 6), providing insight into how LyC clusters contributed to cosmic reionization. Additionally, understanding the relative distribution and geometry of the ISM and ionizing clusters are critical: do we observe density-bounded nebulae or a "picket fence" structure? In other words, how isotropic is the LyC escape? Mapping both neutral and ionized gas features will clarify this geometry and enable us to connect it directly to the underlying stellar populations.


Finally, we need to spatially resolve the feedback-driven structures in and around LyC clusters to better understand the origin of high-pressure interstellar clouds and the mechanisms enabling LyC escape. These clouds are shaped by intense stellar feedback—winds, radiation, and supernovae—that can drive outflows, redistribute metals, and alter the neutral gas distribution in galaxies, all of which are critical to galaxy evolution. Outflows have been linked to LyC escape in complex, time-dependent ways; for example, \citet{Carr25} propose that strong LyC leakers are driven by steady radiative feedback, while weaker leakers are shaped by more chaotic, supernova-driven processes. However, current studies rely on spatially integrated spectra, and existing instruments lack the resolution to examine these mechanisms in detail. Simulations also struggle to resolve cloud disruption, leading to conflicting predictions on the timing and efficiency of LyC escape \citep[e.g.,][]{Trebitsch17, Kimm19, Barrow20}. Recent studies offer complementary perspectives: \cite{Flury22b, Amorin24, Flury25} emphasize the role of ionizing feedback and suggest a duty cycle or bursty star formation as key, while \cite{Bait24} argue that supernovae are not the primary drivers of LyC escape.

Overall, it is critical to have the necessary instruments on HWO to resolve galaxies down to the scales of individual star clusters for the first time. This capability will allow us to directly study how young stars, LyC clusters, and their surrounding environments contribute to the escape of ionizing radiation. By building this foundation, we will finally be able to properly understand the role of galaxies in driving cosmic reionization and shaping the early universe.

\begin{figure*}[h]
\center
	\includegraphics[page = 1,angle=0,trim={0.0cm 0.0cm 0.0cm 0.0cm},clip=true,width=1.0\linewidth,keepaspectratio]{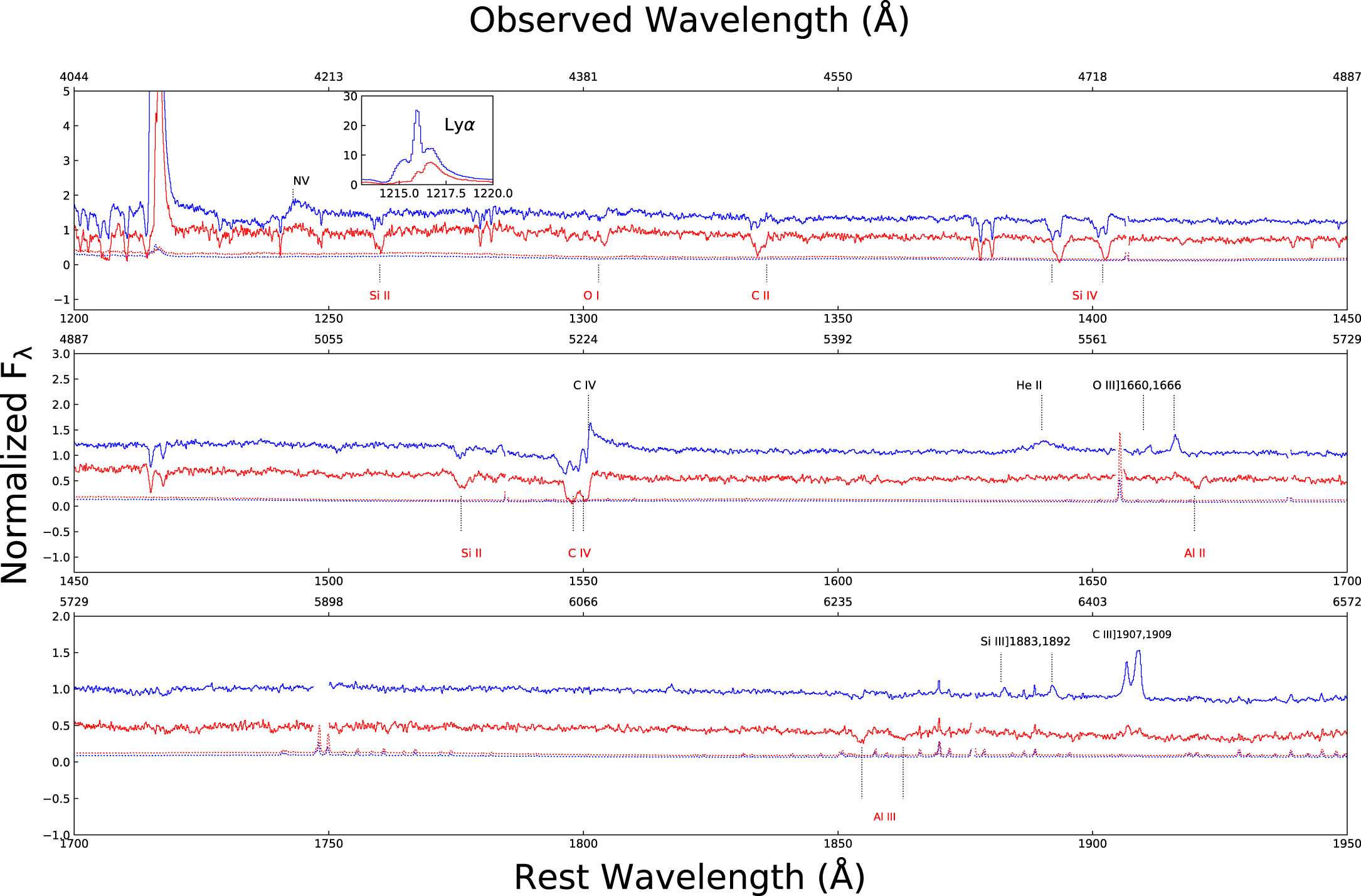}
\caption{\normalfont{Magellan/MagE spectra showing the stacked spectra of the LyC leaking regions (blue) and the non-leaking regions (red) for the gravitational lensed Sunburst arc (z = 2.37). Figure taken from \cite{Mainali22}.}  }
\label{fig:spectra}
\end{figure*}

\section{Physical Parameters}
\label{sec:params}

Based on the science objects, we aim to measure three sets of parameters that are tightly correlated together in Sections \ref{subsec:3.1} -- \ref{subsec:3.3}. Then we describe the detailed requirements for the proposed UV IFU intrument on HWO in Section \ref{subsection:3.4}.

\subsection{Resolve stellar and ionizing radiation in LyC clusters}
\label{subsec:3.1}

We propose to obtain the first large, spatially resolved sample of LyC-emitting star clusters in order to measure their key stellar properties—cluster mass, age, metallicity, dust attenuation, and physical size. These properties will be estimated from cluster-level, resolved SED fits \citep[e.g.,][]{Abdurrouf21, Pascale23}. Stellar metallicity can also be constrained from the P-Cygni stellar wind features found in rest-UV absorption lines, e.g., \civ\ \ly\ly 1548, 1550 and \nv\ \ly\ly 1238, 1240 \citep{Chisholm19}. These properties will be used to derive the total amounts of LyC photons and the shape of ionizing spectra in the clusters.
For the first time in a large sample, we will also directly estimate the LyC escape fraction (\fescLyC) at the level of individual clusters, providing new insight into how ionizing photons propagate through the galaxy and into the IGM. This information is critical, as interpreting galaxy-integrated \fescLyC\ remains highly challenging due to the complex internal structure of galaxies \citep[e.g.,][]{Choi20, Izotov20, Chisholm22, Xu22, Endsley23, Jaskot24a, Jaskot24b, Munoz24, Flury25}. By comparing the distribution of cluster-level \fescLyC to the total escape fraction measured for entire galaxies, we can begin to understand how spatially unresolved measurements relate to the underlying physical conditions—both in the nearby universe and during the epoch of reionization.

\subsection{Resolve the ISM gas around LyC clusters}
\label{subsec:3.2}

We aim to measure the spatially resolved gas-phase abundances around LyC clusters using rest-frame UV emission lines (Figure \ref{fig:spectra}). Key diagnostics include (1) C/O ratio derived from \ciii] \ly\ly 1906, 1908 and [\oiii] \ly\ly 1660, 1666; (2) N/O ratio derived from \niii] \ly\ly 1750, 1752 to [\oiii]; (3) Si/O ratio derived from \siiii] \ly\ly 1883, 1892 to [\oiii] \citep{Pascale23}. 
We will also measure the resolved gas-phase electron density and temperature based on density sensitive line ratios of \ciii] 1907/1909, \siiii] 1883/1892, and \niv\ 1483/1487; and temperature sensitive line ratios of \oiii] 1666/5007 \citep{Mingozzi22}.
Finally, we will measure the resolved nebular dust extinction from rest-optical observations of Balmer emission lines.

\subsection{Resolve cluster’s outflow properties and their feedback effects}
\label{subsec:3.3}

We aim to measure the outflow velocity and full-width-half-maximum (FWHM) of gas from blueshifted UV absorption lines. Key diagnostics include low-ionization lines of \cii\ \ly 1334, \oi\  \ly 1302, and \Siii\ \ly  1260; and high-ionization lines of \siiv\ \ly\ly  1393, 1402 and \civ\ \ly\ly  1548, 1550; along with the Ly-$\alpha$ line \citep[see Figure \ref{fig:spectra},][]{Mainali22}.
We will compare the measured outflow properties with \fescLyC\ escaped from the clusters to understand how outflows impact the escape of ionizing photons \citep[e.g.,][]{Hogarth20, Ferrara24, Carr25}.
In addition, we will compare the measured spatially resolved outflow properties with ISM properties to understand how stellar feedback shapes the geometry and physical state of the interstellar medium.

\subsection{Requirements for proposed UV IFU instrument}
\label{subsection:3.4}

Here we describe the requirements for the proposed UV IFU instrument on HWO to enable robust measurements of the physical parameters described above.

\subsubsection{What are the required spatial resolutions?}

To spatially resolve the LyC clusters, we need resolution down to 10 -- 100 pc. Given the different observational strategies, there are three ways to reach this scale: 1) we can probe the relatively rare gravitational lensed galaxies. In this case, an instrumental spatial resolution of 0.1\arcsec\ could be sufficient if the lensing magnification ($\mu$) is $\sim$ 3 -- 10; 2) we can focus on nearby (z $\ll$ 0.1) SF galaxies, where a spatial resolution of 0.1\arcsec\ can still lead to $<$ 180 pc spaxel$^{-1}$; and 3) the largest breakthrough would be to increase the instrumental spatial resolution to be $\sim$ 0.01\arcsec. In this case, we can conduct large surveys of LyC clusters in SF galaxies up to z $\sim$ 2 ($<$ 80 pc spaxel$^{-1}$). This sample will enable us to fully understand the properties of LyC clusters and their contributions to cosmic reionization.

\subsubsection{What are the required spectral resolutions?}

Current studies of LyC clusters are usually based on the Hubble Space Telescope (HST) and James Webb Space Telescope (JWST), whose spectral resolution is 3000 - 6000, which are sufficient to study the continuum signatures. To spectrally resolve the velocity dispersion of cool clouds in ISM to understand how feedback effects facilitate the escape of LyC radiations, a spectral resolution of $>$~10,000 is required \citep[e.g.,][]{Chen23, Carr25}.

\subsubsection{What is the required wavelength coverage?}

We require the direct detections of LyC, i.e., rest-frame 900 \AA, and the UV continuum and lines, i.e., rest-frame 1000 -- 2000 \AA. The applications of the wavelength coverage constraints to the instrument vary depending on the science samples, and it tradeoffs with the spatial resolution. The lower redshift we probe, the more bluer wavelength coverages and lower spatial resolution are required for the instrument. In additional, we also propose the rest-optical coverage to constrain the ISM properties and geometries, which are also tightly related to LyC escape \cite[e.g.,][]{Xu23, Carr25, Flury25}. This also needs to be done from HWO to reach similar spatial resolution to match them with the UV detections. This will provide a complete view of ionizing photon production and escape mechanisms.

\subsubsection{How deep the observations should be?}

The S/N will be mainly constrained by the faint LyC spectral region. To correctly measure the LyC flux, we require that the S/N is sufficient to detect LyC leakage to $\sim$ 5\% at a S/N = 5, when integrated over a 20\AA\ window around the rest-frame 900\AA. Typical LyC flux for local SF galaxies in this window is $\sim$ 10$^{-19}$ to 10$^{-17}$ ergs s$^{-1}$ cm$^{-2}$ \AA$^{-1}$ $\times$ 20 \AA.

\subsubsection{Summary for the physical parameters}
\label{subsec:summary}

Overall, we summarize all parameters in Table \ref{tab:params} depending on the science sample to be collected by HWO. Specifically, we split the improvements into four stages as follows. 


\begin{enumerate}
    \item State of the Art:  The knowledge of the science question that will be in hand before HWO launches.  Data may exist today, or before HWO launches from space-based or ground observatories, but the future observatory is known to be in work, not just proposed. In our case, there will likely be $\sim$ 10 lensed SF galaxies at z $\sim$ 1 -- 2 and $<$ 100 nearby SF galaxies (z $\ll$ 0.4) with high-fidelity, spatially resolved observations before HWO launches.

    \item Incremental progress (Enhancing):  By changing one or two key observation SOA capabilities in a way that is currently possible, an incremental but non-trivial advance is made over the state of the art. In our case, we expect to have $\sim$ 100 lensed SF galaxies at z $\sim$ 1 -- 2 and $\sim$ 100 nearby galaxies (at least doubling the current sample size) to be observed by HWO with UV IFU instrument to enable the initial investigation of the science goals.

    \item Substantial progress (Enabling):  A significant advance over the state of the art via new capabilities that enables new understanding and enables the right question to be asked to definitively answer the key science question. In our case, we expect to have $\sim$ 1000 unlensed SF galaxies surveyed at z = 0 -- 1.0, which is around 0.1\% of the galaxy samples from Roman and UVEX.

    \item Major progres (Breakthrough): A transformational advance over the state of the art that provides definitive answers to scientific questions, and which is representative of a flagship-class observatory. In our case, we expect to have $\sim$ 10K unlensed SF galaxies surveyed at z = 0 -- 1.0, which is around 1\% of the surveyed ones from Roman and UVEX. 

\end{enumerate}

\begin{table*}[!ht]
  \caption{Observation Strategies and Proposed Parameters to Achieve Different Stages of Science Improvements}
  \label{tab:params}
  \begin{threeparttable}
    \begin{tabular}{%
        p{0.18\textwidth}  
        p{0.17\textwidth}  
        p{0.17\textwidth}  
        p{0.17\textwidth}  
        p{0.17\textwidth}} 
      \tableline
      \noalign{\smallskip}
      Physical Param.\ & State of Art & Incremental Progress & Substantial Progress & Major Progress \\
      \noalign{\smallskip}
      \tableline
      \noalign{\smallskip}
      N of Galaxies$^1$ & $\sim$10 lensed and $<$~100 unlensed & $\sim$100 lensed and 100 unlensed
    &$\sim$1000 unlensed   & $\sim$10,000 unlensed  \\
      \tableline
      \noalign{\smallskip}
      N of SCs$^2$      & $<$ 0.1 -- 10K & $\sim$ 1-10K & $\sim$ 10-100K & $\sim$ 100-1000K \\
      \tableline
      \noalign{\smallskip}
      Spatial Reso. (\arcsec)     & 0.1    & 0.1    & 0.01   & 0.01   \\
      \tableline
      \noalign{\smallskip}
      Spectral Reso. ($R$)        & 3000   & 10,000 & 10,000 & 30,000 \\
      \tableline
      \noalign{\smallskip}
      Wave coverage (\AA)        & 900--2000 & 900--2000 & 900--2000 & 500--2000 \\
      \tableline
      \noalign{\smallskip}
      Sensitivity$^3$ 
                                        & $3\times10^{-18}$ & $1\times10^{-18}$ & $5\times10^{-19}$ & $1\times10^{-19}$ \\
      \noalign{\smallskip}
      \tableline
    \end{tabular}

    \begin{tablenotes}[flushleft]
      \item[$^1$] See the details of lensed and unlensed galaxies in Section \ref{subsec:summary}.
      \item[$^2$] We assume each galaxy has $\sim$ 10 -- 100 LyC clusters to be resolved.
      \item[$^3$] The sensitivity is defined as the average flux density around 900\AA\ rest-frame in unit of erg s$^{-1}$ cm$^{-2}$ \AA$^{-1}$.
    \end{tablenotes}

  \end{threeparttable}
\end{table*}

\section{Description of Observations}
\label{sec:obs}


Detecting LyC radiation is extremely challenging for current state-of-the-art telescopes. First, it requires a space-based observatory, as Earth’s atmosphere absorbs ultraviolet radiation at these wavelengths. Second, it demands high sensitivity in the blue/UV part of the spectrum given the low surface brightness of LyC radiation. Third, direct detections of LyC photons must occur at relatively low redshifts (z $\ll$ 2), because at higher redshifts, absorption by the intervening IGM introduces large uncertainties.

As a result, the ``golden samples" for studying LyC radiation are nearby galaxies, which are apparently brighter and more extended. Currently, the only telescope capable of directly observing their LyC radiation at this redshift is HST with its Cosmic Origins Spectrograph (COS). For instance, HST/COS with the G140L grating can achieve spatially integrated spectra with flux sensitivities down to $3\times10^{18}$ ergs s$^{-1}$ cm$^{-2}$ \AA$^{-1}$ at LyC wavelengths within 2 -- 4 HST orbits. But this is only achievable over a narrow redshift window (z = 0.22–0.38), due to the uneven sensitivity of the detector at the blue end. Moreover, HST expected to decommission in the near future and does not have the capability to resolve the galaxies into small scales. Thus, HWO is essential for continuing and expanding LyC studies into the future and a new UV IFU instrument on it will reach our science goals described in Section \ref{sec:objective}.


The next key question is where we can find a large sample of SF galaxies that produce and allow the escape of large amounts of LyC photons, which are analogs to those at the Reionization Epoch. For lensed SF galaxies, both the Nancy Grace Roman Space Telescope and the Euclid telescope are expected to detect significant numbers across cosmic time \citep{Spergel15, Euclid22, Euclid25}. At redshifts z $<$ 2, Euclid is predicted to discover roughly 100–300 lensed SF galaxies, while Roman is expected to find about 200–500 such systems, thanks to its deeper imaging and higher spatial resolution. Since lensed galaxies will be commonly extended given the lensing magnification, they are ideal candidates for spatially resolved observations. Thus, the most complete LyC survey is to observe all of the lensed SF galaxies from Roman and Euclid with HWO ($\sim$ 300–800 galaxies).

For unlensed SF galaxies, targets will be selected from the millions of low-redshift galaxies detected by current and future ground- and space-based surveys, including Roman’s wide-field imaging programs and UVEX’s FUV and NUV imaging surveys. These rich datasets will provide a large pool of candidates for selecting ideal low-redshift analogs to study LyC radiation with HWO. Among them, we plan a homogeneous search for nearby SF galaxies with extended UV features in images and those show similar spectral features as Green Pea galaxies \citep[e.g.,][]{Henry15}, which are ideal candidates for LyC studies. Thus, we view a substantial to breakthrough project will be covering 0.1\% to 1\% of the targets surveyed by Roman and/or UVEX. Overall, based on these information, we summarize the observation strategies along with the required physical parameters in Table \ref{tab:params}.

{\bf Acknowledgments.} X. Xu acknowledges the fellowship funds provided by Northwestern University's Center for Interdisciplinary Exploration and Research in Astrophysics (CIERA).


\bibliography{author.bib}

\end{document}